\documentclass[preprint,12pt,longnamesfirst]{aastex}

\newcommand \kms          {km~s$^{-1}$}
\newcommand \ha           {H$\alpha$}

\begin{document}

\title{Energy Crisis in the Superbubble DEM\,L\,192 (N\,51D) }
\shorttitle{DEM\,L\,192}

\author{Randall L.\ Cooper\altaffilmark{1,2},  Mart\'{\i}n A.\ 
Guerrero\altaffilmark{1,3},  You-Hua Chu\altaffilmark{1},
C.-H.\ Rosie Chen\altaffilmark{1},  Bryan C.\ Dunne\altaffilmark{1}}
\affil{Astronomy Department, University of Illinois,
1002 West Green Street, Urbana, IL 61801}
\altaffiltext{1}{Astronomy Department, University of Illinois, 
        1002 W. Green Street, Urbana, IL 61801;
        rcooper1@astro.uiuc.edu, mar@astro.uiuc.edu, chu@astro.uiuc.edu
        c-chen@astro.uiuc.edu, carolan@astro.uiuc.edu}
\altaffiltext{2}{Now at Department of Astronomy, Harvard University,
rcooper@cfa.harvard.edu}
\altaffiltext{3}{Now at Instituto de Astrof\'{\i}sica de Andaluc\'{\i}a 
(CSIC), Spain.}



\begin{abstract}

Superbubbles surrounding OB associations provide ideal laboratories 
to study the stellar energy feedback problem because the stellar
energy input can be estimated from the observed stellar content of
the OB associations and the interstellar thermal and kinetic energies
of superbubbles are well-defined and easy to observe.
We have used DEM\,L\,192, also known as N\,51D, to carry out a 
detailed case study of the energy budget in a superbubble, and we find
that the expected amount of stellar mechanical energy injected
into the ISM, $(18\pm5)\times10^{51}$ ergs, exceeds the amount of
thermal and kinetic energies stored in the superbubble,
$(6\pm2)\times10^{51}$ ergs.
Clearly, a significant fraction of the stellar mechanical energy 
must have been converted to other forms of energy.
The X-ray spectrum of the diffuse emission from DEM\,L\,192 requires
a power-law component to explain the featureless emission at 1.0--3.0
keV.  The origin of this power-law component is unclear, but it may be 
responsible for the discrepancy between the stellar energy input 
and the observed interstellar energy in DEM\,L\,192.    


\end{abstract}  



\section{Introduction}

Massive stars inject large amounts of mechanical energy into the ambient
interstellar medium (ISM) via fast stellar winds and supernova explosions.
The stellar energy feedback alters the physical conditions of the ISM 
and consequently may regulate the formation of future-generation stars.
To gain a better understanding of the physical structure and evolution
of the ISM in a galaxy, it is necessary to investigate empirically the
conversion efficiency of stellar mechanical energies into interstellar
kinetic and thermal energies.

Massive stars usually form in groups such as OB associations, and their 
fast stellar winds and supernova ejecta collectively sweep up the ambient
ISM to form superbubbles \citep{MLM88}.  
Superbubbles provide excellent laboratories to study the effects of
stellar energy feedback on the ISM because the kinetic and thermal
energies in a superbubble are well-defined and can be measured with
little ambiguity.
Superbubbles in the Large Magellanic Cloud (LMC) at 50 kpc  are 
particularly well suited for such studies, as their stellar content and 
ISM can be observed in great detail with little foreground extinction.

We have chosen the superbubble DEM\,L\,192 \citep{DEM76} in the LMC
for a detailed study of the interaction between its stars and the
ambient ISM.
DEM\,L\,192, also known as N\,51D \citep{Henize56}, hosts two OB
associations, LH\,51 and LH\,54 \citep{LH70}.
The stellar content of these two OB associations and the dynamics
of the superbubble have been analyzed in detail by \citet{Oey96a} and
\citet{OS98}.
They conclude that LH\,51 and LH\,54 are both young,
$\sim3\times10^6$ yr old.
The diffuse X-ray emission from the hot interior of DEM\,L\,192
was detected by {\sl Einstein} and {\sl ROSAT} observations 
\citep{CM90,DPC01}, but high-quality spectra of the X-ray emission 
have been provided only by {\sl XMM-Newton} observations.
The emission line features in the X-ray spectra suggest an 
overabundance in oxygen and neon which is used to argue for
the occurrence of supernova explosions \citep{Betal02}.

This paper reports our investigation of the energetics of the 
superbubble DEM\,L\,192.
In \S 2 we analyze archival {\sl XMM-Newton} observations of 
DEM\,L\,192 to determine the physical properties of the hot gas 
in the superbubble interior.
In \S 3 we determine the thermal energy of the hot superbubble
interior and the kinetic energies of the ionized and neutral gas 
in the superbubble shell.
In \S 4 we compute the stellar wind energy and supernova explosion energy 
expected from the known stellar content of LH\,51 and LH\,54.
Finally, a discussion of the overall energy budget of the 
DEM\,L\,192 superbubble and our conclusions are given in \S 5.

\section{XMM-Newton Observations of Hot Gas in DEM\,L\,192} 

DEM\,L\,192 was observed with the {\sl XMM-Newton Observatory} on 
2001 October 31 and November 1 using the EPIC/MOS1, EPIC/MOS2, and 
EPIC/pn CCD cameras (Obs.\ ID: 7194).  The two EPIC/MOS 
cameras were operated in the Prime Full-Window Mode for a useful exposure 
time of 31.6 ks, and the EPIC/pn camera was operated in the Extended Prime 
Full-Window Mode for a useful exposure time of 24.6 ks.  The thin filter was 
used for all observations.  

We retrieved the {\sl XMM-Newton} pipeline products from the {\sl XMM-Newton} 
Science Archive (XSA)\footnote{
The {\sl XMM-Newton} Science Archive is supported by ESA and can be accessed 
at http://xmm.vilspa.esa.es/xsa.} 
and further processed the data using the {\sl XMM-Newton} Science Analysis 
Software (SAS ver.\ 5.3.3) and the calibration files from the 
Calibration Access Layer available on 2002 December 13.  
The event files were screened to eliminate bad events (e.g., events due 
to charged particles) so that only events with CCD patterns 0-12 
(similar to {\sl ASCA} grades 0-4) were selected for the EPIC/MOS 
observations, and events with CCD pattern 0 (single pixel events) were
selected for the EPIC/pn observation.  
We also screened the event files to remove periods of high background.  
To assess the background rate, we binned the counts over 50 s time 
intervals for each instrument in the 10-12 keV energy range which is 
dominated by the background.  
Only for a short period of time did the high background, with count rates 
$\geq$0.3 counts~s$^{-1}$ for the EPIC/MOS or $\geq$0.6 counts~s$^{-1}$ 
for the EPIC/pn, affect the observations; events from this high-background 
period were discarded.  The resulting net exposure times are 30.6 ks and 
23.2 ks for the EPIC/MOS and EPIC/pn observations, respectively.  

\subsection{Spatial Analysis}

We extracted images from the EPIC/pn and EPIC/MOS observations in the 
0.25--2.5 keV energy band, which includes most of the diffuse X-ray 
emission from DEM\,L\,192. 
These images were then combined to obtain a higher S/N ratio image and 
to reduce the null exposure in the gaps between CCDs (the satellite was 
not dithered during the observations).  The resulting raw EPIC image is 
presented in Fig.~1a.  This image was further adaptively smoothed using 
Gaussian profiles with FWHM ranging from 2\arcsec\ to 15\arcsec\ and divided 
by a normalized exposure map to remove the vignetting of the telescope and 
sensitivity variations of the instruments.  A grey-scale presentation of 
this adaptively smoothed EPIC image is shown in Fig.~1b.  For comparisons,
an H$\alpha$ image of the same region is displayed in Fig.~1c, and the 
X-ray contours extracted from the smoothed EPIC image are overplotted 
on the H$\alpha$ image in Fig.~1d.

Diffuse X-ray emission from the interior of DEM\,L\,192 is clearly detected.
The brightest X-ray emission region is coincident with the OB association 
LH\,54, peaking particularly at the bright Wolf-Rayet star HD\,36402.
The X-ray peaks coincident with bright stars may be dominated by stellar 
emission; therefore, emission from these peaks is excluded from our 
analysis of diffuse X-rays.
The diffuse X-ray emission shows a rough limb-brightened morphology in the
main body of the superbubble, with the brightest diffuse X-ray emission
projected interior to the brightest H$\alpha$ emission regions of the shell.
Faint diffuse X-ray emission is detected to the north and southwest of the 
main body; however, these faint X-ray extensions do not necessarily imply 
blowouts, as they are bounded by faint H$\alpha$ filaments externally.
The superbubble DEM\,L\,192 must be in an inhomogeneous ISM, and the 
variations in the H$\alpha$ surface brightness reflect the density 
variations in the ambient ISM.
The rough correspondence between the diffuse X-ray surface brightness and 
the nebular H$\alpha$ surface brightness suggests that a mixing of the 
dense, cooler, superbubble shell gas into the hot interior has occurred.

\subsection{Spectral Analysis}

The X-ray spectrum of DEM\,L\,192 is extracted from only the event file 
of the EPIC/pn camera because its sensitivity is greater than that of the 
EPIC/MOS cameras.  The source aperture, marked in Fig.~1a, includes a
large elliptical region that covers the main body of DEM\,L\,192 and the 
north extension and a smaller square region that covers the southwest 
extension; these two regions together encompass all diffuse X-ray emission
within the superbubble.  The background contribution 
was assessed using an aperture exterior to DEM\,L\,192 with similar area to 
the source aperture.  All point sources were excised from the source and 
background apertures.  A circular region that includes most stars in 
LH\,54 is also excluded from the source aperture to minimize the stellar 
contribution to diffuse emission.   After the background subtraction, a 
total of $\sim$16,000 counts are detected in the source aperture.  
Some residuals of the background subtraction are noticeable at the 
energy of the Al K$\alpha$ fluoresce line at $\sim$1.5 keV.  
The energy range 1.45-1.55 keV, thus, has been omitted in the subsequent 
spectral analysis.  

The background-subtracted EPIC/pn spectrum of DEM\,L\,192, shown in Fig.~2,
peaks at 0.5--0.6 keV, with a bright shoulder at $\sim$0.7 keV and a weaker 
peak at 0.9--1.0 keV.  Above 1.0 keV, the spectrum is featureless with 
emission detected up to 3.5 keV.  
The overall spectral shape indicates thermal plasma emission, with 
spectral features corresponding to the He-like triplet of O~{\sc vii} 
at $\sim$0.57 keV, the H-like O~{\sc viii} line at $\sim$0.65 keV, 
and the He-like triplet of Ne~{\sc ix} at $\sim$0.92 keV.  
Therefore, we have adopted the MEKAL optically-thin plasma emission model 
\citep{KM93,LOG95} with 1/3 $Z_\odot$ abundances as appropriate for the 
LMC and the absorption cross-sections from \citet{BM92}.  
The spectral fits are carried out using XSPEC ver.\ 11.2.0 by folding 
the model spectrum through the EPIC/pn response matrices and comparing 
the modeled spectrum to the observed spectrum in the 0.3--4.5 keV 
energy range.
The $\chi^2$ statistics are used to determine the best-fit model.  

The best-fit MEKAL model with 1/3 $Z_\odot$ abundances is very 
unsatisfactory (reduced $\chi^2$ = 7.2), 
failing to fit the spectrum at 1--3 keV and showing large residuals at 
the Ne~{\sc ix} lines at $\sim$0.92 keV.  
Adding a second thermal component at a higher temperature can improve
the spectral fits significantly (reduced $\chi^2$ = 2.2), but the 
temperature of the second component is pegged at the upper limit 
allowed by XSPEC, i.e., $kT$ = 80 keV.
A thermal component at such a high temperature is not expected in
the ISM; furthermore, it is dominated by bremsstrahlung emission
of which the spectral shape is featureless and similar to that of a power law.
Therefore, we have added a power-law component, instead of a second
thermal component, and the reduced $\chi^2$ is lowered to 2.2 for
the best-fit model.
The large residuals at the Ne~{\sc ix} lines remain, and they 
can be minimized only if the Ne abundance is allowed to vary.
By increasing the Ne abundance by a factor of 3 (i.e., solar Ne
abundance and 1/3 solar for the other elements), we are able to
obtain a best-fit model with a reduced $\chi^2$ of 1.4, as shown
in Fig.~2.
This confirms the enhanced Ne abundance suggested by \citet{Betal02};
however, we do not find compelling evidence for an enhanced O abundance,
which has also been suggested by \citet{Betal02}.

In our best-fit model for the diffuse X-rays from DEM\,L\,192,
the thermal component dominates the emission below 1 keV with a
temperature $kT$ of 0.21$\pm$0.01 keV, an absorption column density of
(3.4$\pm$1.7)$\times$10$^{20}$ H-atom cm$^{-2}$, and a normalization 
factor\footnote{The normalization factor $A$ is defined to be 
$A$=$1\times10^{-14}{\int}N_{\rm e}^2 \mathrm{d}V/{4{\pi}d^2}$,
where $d$ is the distance, $N_{\rm e}$ is the electron number density,
and $V$ is the volume.}  
$A = (1.6\pm0.4) \times 10^{-3}$ cm$^{-5}$.  
The power law component dominates the emission from 1 to 3 keV
and has a photon index $\Gamma$ of $1.3 \pm 0.2$. 
The observed flux of diffuse X-ray emission from DEM\,L\,192 is 
$1.1 \times 10^{-12}$ ergs~cm$^{-2}$~s$^{-1}$ and the X-ray luminosity is 
$\sim 4 \times 10^{35}$ ergs~s$^{-1}$ in the 0.3--3.0 keV band. 
The thermal emission accounts for $\sim$70\% of the observed luminosity 
(but $\sim$85\% of the total number of detected photons).  

\section{Energy Content of DEM\,L\,192}

For a superbubble like DEM\,L\,192, we expect the stellar mechanical energy
injected into the ISM to be stored in the thermal energy of its hot interior 
and the kinetic energy of the dense, swept-up shell.  As the ionization front
might be trapped in the swept-up shell, both ionized and neutral gas in the
superbubble shell must be considered.  Below we examine separately the 
thermal energy of the hot interior and the kinetic energies of the ionized 
shell gas and associated H~{\sc i} gas. 

\subsection{Thermal Energy of the Superbubble Interior}

The thermal energy of the hot gas in a superbubble is
$ E_{\rm th} = (3/2) k T N \epsilon_{1} V$,
where $T$ is the plasma temperature, $N$ is the total particle number 
density, $\epsilon_{1}$ is
the volume filling factor of the hot gas, and $V$ is the volume interior to 
the superbubble.
Assuming a canonical helium to hydrogen number density ratio of
$N_{\rm He}/N_{\rm H}$ = 0.1, the total particle number density, 
approximated by the sum of electron density, hydrogen density, and 
helium density, is 
$N \sim N_{\rm e}+N_{\rm H}+N_{\rm He}~\sim~1.92 N_{\rm e}$.
For a given volume filling factor $\epsilon_{1}$, the rms $N_{\rm e}$ 
can be derived from the normalization factor $A$ of the best-fit model: 
$N_{\rm e}$ = $2 \times 10^{7}d(\pi A/\epsilon_1 V)^{1/2}$.
For a distance of $d$ = 50 kpc, a normalization factor of 
$A = (1.4 \pm 0.4) \times 10^{-3}$~cm$^{-5}$, and a volume 
$V = 4.4\times10^{61}$ cm$^3$ (approximated by a 
$120\times120\times200$ pc$^3$ ellipsoid), we find 
$N_{\rm e} = (0.031 \pm 0.004)~\epsilon_{1}^{-1/2}$ cm$^{-3}$. 
The total mass of the hot gas in the superbubble is
$1.3\times10^3~\epsilon_1^{1/2}$ M$_\odot$, and the total thermal energy 
of the hot gas is $E_{\rm th} = 
(1.3 \pm 0.3) \times 10^{51}~\epsilon_{1}^{1/2}$ ergs. 
The filling factor $\epsilon_{1}$ is most likely between 0.5 and 1.0. 
For $\epsilon_{1} = 0.5$, $N_{\rm e}$ = $0.044 \pm 0.004$ cm$^{-3}$,
the hot gas mass is 920 M$_\odot$,
and $E_{\rm th} = (9 \pm 2) \times 10^{50}$ ergs.

The cooling timescale of the hot gas
$t_{\rm cool}$ is $ \sim {(3/2) N k T}/\Lambda(T)$,
where $\Lambda(T)$ is the cooling function for gas at temperature $T$
in units of ergs cm$^{-3}$ s$^{-1}$.
For a plasma temperature of a few $\times$ 10$^{6}$ K, 
$\Lambda(T)/N_{\rm H}^2 = 7 \times 10^{-24}$ ergs cm$^{3}$ s$^{-1}$ 
for 1/3 Z$_\odot$ abundances \citep{DM72}.  
With $N_{\rm H} \sim 0.83 N_{\rm e}$, $N \sim 1.92 N_{\rm e}$, 
and $kT = 0.21 \pm 0.01$ keV, the cooling timescale $t_{\rm cool}$
is $\sim 200$ Myr. 
Since the cooling timescale is much larger than the age of the 
OB associations LH\,51 and LH\,54, $\sim$ 3 Myr, only a negligible 
amount of thermal energy has been radiated away.

\subsection{Kinetic Energy of the Ionized Superbubble Shell}

The mass of the ionized superbubble shell can be derived from the 
H${\alpha}$ luminosity, $L_{\rm H{\alpha}}$.
The H${\alpha}$ flux from DEM\,L\,192 is $f_{\rm H{\alpha}} = 
(860 \pm 80) \times 10^{-12}$ 
ergs s$^{-1}$ cm$^{-2}$ \citep{KH86}.  For a distance of
$d = 50$ kpc, $L_{\rm H{\alpha}} = 4\pi d^{2} f_{\rm H{\alpha}} = 
(2.6\pm0.3) \times 10^{38}$ ergs s$^{-1}$.  
In addition, for ionized gas at temperatures of $\sim$10$^4$ K, 
$L_{\rm H{\alpha}} = 3.56 \times 10^{-25} N_{\rm e}N_{\rm p}V\epsilon_{2}$ 
ergs s$^{-1}$, where $N_{\rm e}$ and $N_{\rm p}$ are the particle number 
densities of electrons and protons in units of cm$^{-3}$, respectively, 
$V$ is the volume of the region in units of cm$^{3}$, and $\epsilon_{2}$ is 
the volume filling factor.  
Again, we approximate the superbubble volume by a $120\times120\times200$ 
pc$^3$ ellipsoid, and the volume is $V = 4.4\times10^{61}$ cm$^3$.
Helium is most likely singly ionized, thus
$N_{\rm e} = 1.1 N_{\rm p} = (4.5\pm0.5)~\epsilon_{2}^{-1/2}$ cm$^{-3}$.
The total mass of the superbubble shell is $M = 
1.27 N_{\rm e} V \epsilon_{2} m_{\rm H}$, where $m_{\rm H}$ is the 
mass of a hydrogen atom.
Using the rms $N_{\rm e}$ derived from $L_{\rm H{\alpha}}$, we find
the superbubble shell mass to be
$M = (4.2\pm0.5)\times10^{38}~\epsilon_{2}^{1/2}$ g = 
$(2.1\pm0.2)\times10^5~\epsilon_{2}^{1/2}$ M$_\odot$.
The expansion velocity of the DEM\,L\,192 superbubble has been
measured in the H$\alpha$ line to be $v_{\rm exp} \sim 35$ 
km~s$^{-1}$ \citep{MT80}.
Thus, the kinetic energy in the ionized superbubble shell is
$E_{\rm kin} = \frac{1}{2} M v_{\rm exp}^{2}$ = 
$(2.6 \pm 0.3) \times 10^{51}\epsilon_{2}^{1/2}$ ergs.
If the thickness of the ionized superbubble shell is 1/10 of the 
radius, then the filling factor of shell gas is $\epsilon_{2} = 0.27$,
the ionized shell mass is $M \sim 1\times10^5$ M$_\odot$, and the 
ionized shell kinetic energy is $E_{\rm kin} \sim 1\times10^{51}$ ergs.

\subsection{Kinetic Energy of Associated H~{\sc i} Gas}

The kinetic energy of the neutral gas in the swept-up shell can 
be derived from observations of the 21-cm line emission from the 
\ion{H}{1} gas.  Such observations of DEM\,L\,192 have been extracted 
from \ion{H}{1} surveys of the LMC made with the Australia Telescope 
Compact Array (ATCA) and the Parkes Observatory 64-m telescope, where 
single-dish observations from the Parkes telescope have been
combined with the ATCA data to ``fill-in'' low spatial frequencies 
not observed by the interferometer.  The ATCA and Parkes observations 
are detailed in \citet{Kim98} and \citet{Kim03}, respectively.

The kinematics of the \ion{H}{1} gas can be determined from 
the position-velocity plots in Fig.~3.
The \ion{H}{1} gas shows a broad velocity profile centered at the 
systemic velocity of $V_{\rm hel} \sim 300$~\kms~and with a width of 
$\sim$70~\kms, corresponding to twice the expansion velocity measured 
in \ha\ emission, but no clear expanding shell structure can
be identified.
The \ion{H}{1} channel maps \citep{DCS04} and interstellar absorption 
observations of stars in DEM\,L\,192 \citep{Wetal04} indicate that the 
\ion{H}{1} velocity structure is rather complex and some velocity 
components may not be physically associated with the superbubble.

A total 21-cm line flux density of 470~Jy is measured from the combined 
radio observations within the dotted polygon outlined in the \ion{H}{1}
column density map in Fig.~3.
Given the LMC distance of 50~kpc, we estimate 
a total \ion{H}{1} mass of 3~$\times$~10$^5$~M$_\odot$ from this flux 
density.  The total kinetic energy of the \ion{H}{1} gas can be 
approximated by 3/2 of the second moment of the integrated velocity 
profile, assuming that the motion is isotropic. 
This estimated kinetic energy is $\sim3\times10^{51}$ ergs.
We can also determine the \ion{H}{1} kinetic energy using the total 
\ion{H}{1} mass and the expansion velocity of the superbubble shell; 
the resulting kinetic energy is $3.7\times10^{51}$ ergs, compatible 
with the \ion{H}{1} kinetic energy estimated from the second moment 
of the velocity profile.
Note that these \ion{H}{1} mass and kinetic energy estimates should 
be considered as upper limits, as some of the \ion{H}{1} gas may
not be physically associated with the superbubble.

\section{Stellar Energy Injected into DEM\,L\,192}

Stellar mechanical energy is injected into the ambient ISM via fast
stellar winds and supernova explosions.  
Below we determine the contributions of these two types of stellar 
energy feedback.

\subsection{Stellar Wind Energy}

To determine the stellar wind energy injected into DEM\,L\,192, we 
consider the 50 most luminous stars in the OB associations LH 51 
and LH 54 given by \citet{OS98}.
The spectroscopically determined spectral types of these stars are 
matched with the theoretical models of \citet{SdK97} to assess the
stellar mass $M$, mass loss rate $\dot{M}$, and stellar wind terminal 
velocity $v_{\infty}$.
Since $Z_{\rm LMC} \sim 1/3~Z_{\odot}$, we apply the necessary 
corrections to the listed theoretical values of $\dot{M}$ and $v_{\infty}$ 
for low metallicity stars as described in \citet{SdK97}.  
From these data, we estimate the stellar wind luminosity by $L_{\rm w} = 
(1/2)\dot{M}v_{\infty}^{2}$.  The metallicity-corrected parameters and 
mechanical luminosities of the 50 stars are listed in Table 1 in the same 
order as given in Table 1 of \citet{OS98}.
We calculate the total mass loss rate to be $4.4\times10^{-5}$ 
M$_\odot$~yr$^{-1}$ and the total stellar wind luminosity to be 
$5.5 \times 10^{37}$ ergs s$^{-1}$.  
Thus, for an age of $\sim 3$ Myr \citep{OS98}, the total stellar mass
loss is $\sim100$ M$_\odot$, and the total wind energy input is about 
(5 $\pm$ 1) $\times$ 10$^{51}$ ergs.

\subsection{Supernova Energy}

To estimate the number of stars that have already exploded as supernovae, 
we use the combined mass function of LH\,51 and LH\,54.
$UBV$ photometry of 456 stars in LH\,51 and LH\,54 is available from
\citet{Oey96a}.
The mass of each main sequence star is determined by comparing its 
location on a color-magnitude diagram (CMD) to the evolutionary tracks 
of stars with different masses.
As the reddening is not known for individual stars, we use the
extinction-free magnitude $W = V - DM - (B-V) A_V/ E(B-V)$ 
and reddening-free color $Q = (U-B) - (B-V) E(U-B) / E(B-V)$,
where $DM$ = 18.5 is the distance modulus of the LMC,
${E(U-B)}/{E(B-V)} = 0.72$, and ${A_V}/{E(B-V)} = 3.1$.
Fig.~4 shows the reddening-free CMD of LH\,51 and LH\,54, where the
evolutionary tracks of main sequence stars with $Z = 0.4~Z_{\odot}$
provided by \citet{LS01} are plotted for initial masses of 5, 7, 10, 12,
15, 20, 25, 40, and 60 M$_\odot$.
Stars in each mass bin are counted.
We assume a Salpeter initial mass function for LH\,51 and LH\,54:
$f(M) = KM^{-2.35}$, where $K$ is a constant that can be determined
from the star counts.
The number of stars with masses between M$_{\rm min}$ and M$_{\rm max}$
is $N = \int_{{\rm M}_{\rm min}}^{{\rm M}_{\rm max}} f(M)\,\mathrm{d}M$.
For a M$_{\rm min}$ of 5--7 M$_\odot$ and a M$_{\rm max}$ of 
20--25 M$_\odot$, we find K = 3040 $\pm$ 460.
This constant is insensitive to the choice of M$_{\rm max}$ but 
rather sensitive to the choice of M$_{\rm min}$; the error bar reflects 
the uncertainty caused by different M$_{\rm min}$.
The number of stars with masses greater than 25 M$_{\odot}$ is
$\int_{25 {\rm M}_\odot}^{100 {\rm M}_\odot} 
f(M)\,\mathrm{d}M = 25 \pm 4$.
From Table 1, we find 12 stars with masses greater than 25 M$_{\odot}$, 
so the number of stars that have exploded in supernovae is 
$(25 \pm 4) - 12 = 13 \pm 4$.
We have chosen to count main-sequence stars in the 5--25 M$_{\odot}$ 
range because the evolutionary tracks of such stars are well constrained
by the empirically derived relationship between stellar masses and spectral 
types and are not subject to the uncertainties for the most massive
stars \citep[e.g.,][]{Aetal99}.
We have compared the evolutionary tracks of stars in the 5--25 M$_{\odot}$ 
range between the Geneva models \citep{SMS93} and the Padua 
models \citep{Fetal94}, and find that they are indeed similar.
Assuming that each supernova releases about 10$^{51}$ ergs of explosion
energy, the total supernova energy input is approximately 
(1.3 $\pm$ 0.4) $\times$ 10$^{52}$ ergs.

\section{Energy Budget of the Superbubble DEM\,L\,192}

The energies stored in the superbubble DEM\,L\,192 and the stellar
energy input are summarized in Table 2.
The values and error bars of the thermal and kinetic energies
take into account the most probable range of filling factors.
The stellar energy input includes contributions from fast
stellar winds and supernovae.
The superbubble energy includes the thermal energy of its hot 
interior and the kinetic energy of the superbubble 
shell.\footnote{The thermal energy of the ionized superbubble shell
is not included in the energy budget discussion because the
superbubble shell is photoionized and the thermal energy is
provided by the UV radiation of the OB associations.}
It is immediately clear that the expected amount of stellar energy 
injected into the ISM is $\sim3$ times as large as that observed 
in the superbubble. 
 
Previously, observed dynamics of bubbles blown by single stars or
superbubbles have been compared with pressure-driven bubble models
by \citet{Wetal77} or variations of such models, and it has been
commonly concluded that the observed stellar wind strength is too 
high for the observed bubble dynamics or that the bubble/superbubble
shell expands too slowly for the given stellar wind strengths
\citep[e.g.,][]{GLM96,Oey96b,Netal01}.
These results are equivalent to what we have found in DEM\,L\,192,
i.e., the stellar mechanical energy injected into the ISM
exceeds the energies storied in the superbubble.
More specifically, according to the \citet{Wetal77} model, the 35 \kms\
expansion velocity and 60 pc radius of DEM\,L\,192 indicate a 
dynamic age of 1 Myr, which is shorter than the age of the OB 
associations LH\,51 and LH\,54. 
The mass of the ionized superbubble shell consists of the material 
swept up from the superbubble interior, thus the volume and mass 
of the \ion{H}{2} gas given in \S3.2 imply an ambient density of 
$\sim 3$ H-atom cm$^{-3}$.
The wind mechanical luminosity required by \citet{Wetal77} for 
DEM\,L\,192 is $1.5\times10^{38}$ ergs~s$^{-1}$, and the total 
energy required during the 1 Myr dynamic time is 
$\sim5\times10^{51}$ ergs, which is much smaller than the total 
stellar energy expected from LH\,51 and LH\,54.

Evidently there is an energy crisis: the observed interstellar
energies cannot account for all the stellar energy injected into 
the ISM.  Several mechanisms may drain energies from a superbubble. 
The most obvious mechanism is a superbubble blowout or breakout, 
in which the interior hot gas spews out, directly reducing the
thermal energy and pressure of the superbubble interior.
However, no blowouts are obvious in DEM\,L\,192.  As shown in 
Fig.~1, the extended diffuse X-ray emission from DEM\,L\,192 is 
completely bounded by outer, curved \ha\ filaments, in sharp 
contrast to the blowout of the superbubble DEM\,L\,152, also
known as N44, in which the diffuse X-ray emission extends beyond 
open, streamer-like \ha\ filaments \citep{Cetal93}.

Another mechanism that may drain the internal energy of a superbubble
is the evaporation of dense cooler gas into the hot interior.
In addition to the dense ionized superbubble shell, DEM\,L\,192 
contains dense cloudlets with ionized surface layers, as revealed 
in high-resolution {\sl HST} WFPC2 \ha\ images \citep{Cetal00}.
The main energy loss would occur at the interface layer between the
$10^6$ K hot gas and the $10^4$ K cool gas because the cooling rate 
is the highest for temperatures of a few $\times 10^5$ K.
We may estimate the energy loss due to cooling from interface layers.
Assuming that the interface layers are in pressure equilibrium
with the hot interior gas, their electron number density would be 
0.3--0.4 cm$^{-3}$.
The total volume of the interface layers is nevertheless unknown 
because no interface layer has ever been spatially resolved 
observationally and even the existence of heat conduction and
mass evaporation may be questionable \citep{RS03}.
The spatial extent of an interface layer can be estimated from 
the relative locations of diffuse X-ray emission and the optical
shell of a bubble or superbubble.
The highest-quality X-ray observations of wind-blown bubbles are
the {\sl Chandra} observation of NGC\,6888 and the {\sl XMM-Newton}
observations of S308 \citep{Getal04,Cetal03}.
While NGC\,6888 shows no appreciable separation between the outer 
edge of the diffuse X-ray emission and the outer edge of the optical
shell, S308 shows a gap of 90--200$''$ (or 0.5 to 1.7 pc), which 
contains the dense ionized gas shell and the interface layer.
The interface layer must be less than 0.5 pc in thickness.
Adopting this upper limit of the interface layer thickness for 
DEM\,L\,192, the volume occupied by the interface layer is no
more than 2.5\% of the total volume.
Using a cooling function $\Lambda(10^5 {\rm K})/N_{\rm H}^2 \sim 3 
\times 10^{-22}$ ergs cm$^3$ s$^{-1}$ appropriate for the
LMC metallicity \citep{DM72}, the current cooling rate would be
$4\times10^{37}$ ergs~s$^{-1}$. 
If the superbubble has been evolving with the radius $\propto t^\eta$,
where $t$ is the age and $\eta$ is a constant (e.g., $\eta$ = 0.6 for
the Weaver et al.\ model), then the volume has been growing 
$\propto t^{3\eta}$, and the total energy radiated away in 3 Myr by 
the interface layer would be $4\times10^{51} (3\eta+1)^{-1}$ ergs.
If $\eta$ = 0.6, the total radiative loss is 
$\sim1.5\times10^{51}$ ergs.
This generous upper limit for radiative loss is non-negligible but not
sufficiently significant in the energy budget.
Therefore, radiative loss in the interface layer between dense cold gas
and the hot superbubble interior cannot be the cause of the energy 
discrepancy.

Where did the stellar energy go if it is not converted into the
thermal and kinetic energies of the ISM? 
The interaction between a fast moving plasma, being it tenuous
stellar winds or dense supernova ejecta, and the ambient ISM
is a complex magnetohydrodynamic problem \citep{Sp78}.
It is perhaps an oversimplification to assume that supernova
ejecta and fast stellar winds are thermalized in the hot
interior of a superbubble.
In fact, the {\sl XMM-Newton} X-ray observations of DEM\,L\,192
reveal a possible mechanism that may play an important role in 
the energy budget.
Our analysis of the X-ray spectrum of the diffuse emission from
DEM\,L\,192 indicates that not all emission originates from a
thermal plasma; a power-law component is needed to explain the
emission in the 1.0--3.0 keV range.
This power-law component cannot originate from unresolved
point sources, as the smoothed X-ray image in the 1.1--3.5 keV 
band displayed in Fig.~5 shows that this hard X-ray emission 
is diffuse.
It is possible that this power-law component originates from the interactions 
of supernova ejecta and fast stellar winds with the hot interior of the 
superbubble, but the exact origin is unclear.
It is unlikely that this power-law component represents a synchrotron 
X-ray emission similar to those seen in young supernova remnants such as 
SN 1006 \citep{1995Natur.378..255K} and Cas A \citep{2003ApJ...584..758V}
and the superbubble 30 Dor C \citep{bamba} for two reasons.  First,  the 
X-ray photon index of DEM\,L\,192, $\Gamma = 1.3\pm0.2$, is lower than those 
observed in the other objects, $\Gamma \sim 2$.  Second, the synchrotron 
X-ray emissions from young supernova remnants and 30 Dor C show 
limb-brightened morphologies because the relativistic particles are 
accelerated 
in the shocks, but the power-law X-ray emission from DEM\,L\,192 does not 
show enhancement along the edge (see Fig.\ 5).
Observations of more superbubbles are needed to determine whether
a power-law component is commonly needed to model the spectra of
their diffuse X-ray emission.
Future theoretical models of superbubbles need to solve
the plasma interaction problem realistically instead of making
simplifying assumptions about shocks and converting mechanical
energy into only kinetic and thermal energies, so that true
progress in the understanding of bubbles and superbubbles can be
made.

\acknowledgments
YHC acknowledges the support of NASA grant NAG 5-13076.  We thank the referee 
for prompt reading and providing useful suggestions to improve this paper.
We also thank Ya\"el Naz\'e for her critical reading of the paper.

\newpage

\begin{figure}

\plotone{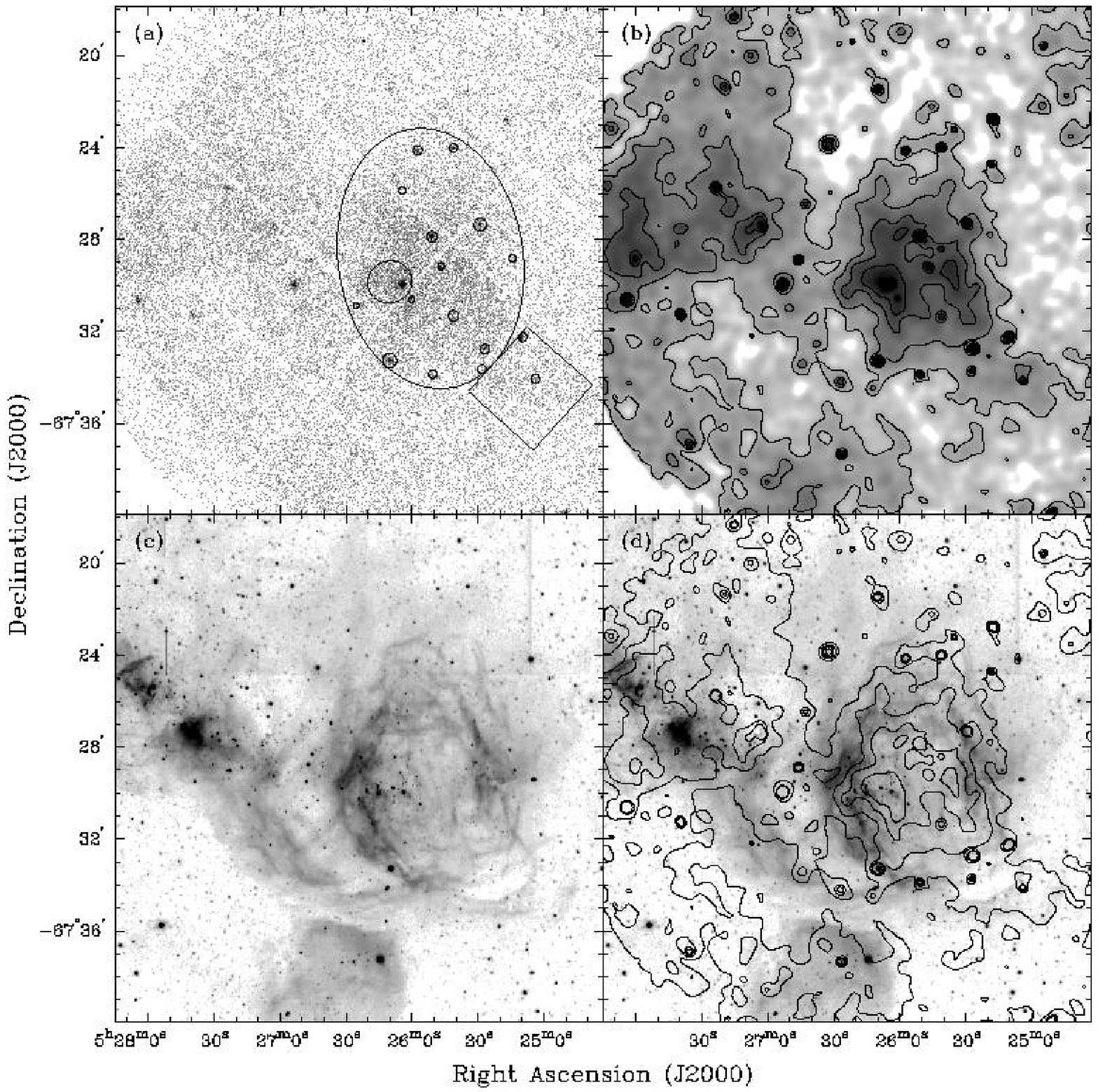}
\caption{(a) {\sl XMM-Newton} EPIC image of DEM\,L\,192.
The spectrum of the diffuse X-ray emission is extracted from
the source aperture consisting of elliptical and square regions
marked.  The point sources, marked by small circles, and
stellar emission from the OB association LH\,54, marked by
a larger circle, are excluded from the source aperture.
(b) Adaptively smoothed and vignetting-corrected X-ray image.
The contour levels are at 9, 15, 22.5, and 30 $\sigma$ above 
the background. 
(c) \ha\ image of  DEM\,L\,192.
(d) \ha\ image overplotted with X-ray contours.
}
\label{fig1}
\end{figure}

\begin{figure}
\plotone{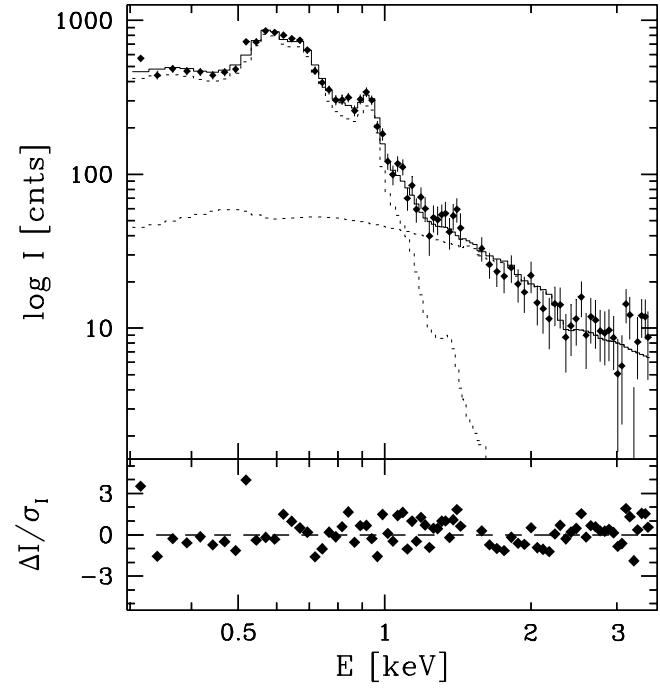}
\caption{{\sl XMM-Newton} EPIC/pn spectrum of the diffuse
X-ray emission from DEM\,L\,192.  The best-fit model is overplotted
in the solid curve.  The thermal plasma component and the power-law 
component are individually plotted in dashed curves.  The bottom
panel plots the residuals of the fit.
}
\label{fig2}
\end{figure}

\begin{figure}
\begin{center}
\includegraphics[angle=270,width=0.95\textwidth]{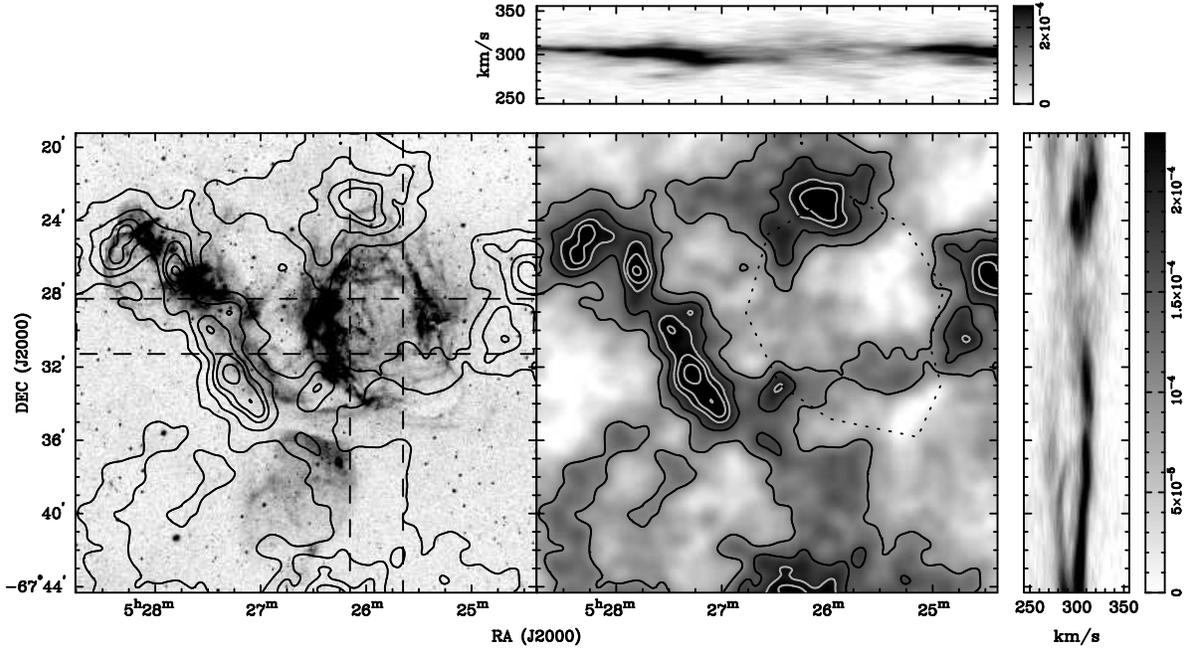}
\end{center}
\caption{H$\alpha$ image (left) and \ion{H}{1} 21-cm line map (right) 
of DEM\,L\,192.  Both images are overlaid with \ion{H}{1} 21-cm line 
emission contours at levels of 50\%, 65\%, 77\%, 85\%, and 95\% of 
the peak.  The dotted polygon plotted over the \ion{H}{1} map represents 
the extent of the superbubble and the region over which the \ion{H}{1}
mass is extracted.
We have produced position-velocity plots for two 3$'$-wide slices 
through the superbubble along the right ascension and declination axes.  
These slices are shown above and to the right, respectively, of the 
\ion{H}{1} image accompanied with greyscale wedges in arbitrary units.  
The velocities are heliocentric.  The area covered by each slice is 
indicated by the dashed lines on the \ha\ image.
}
\label{figHI}
\end{figure}

\begin{figure}
\epsscale{1.0}
\plotone{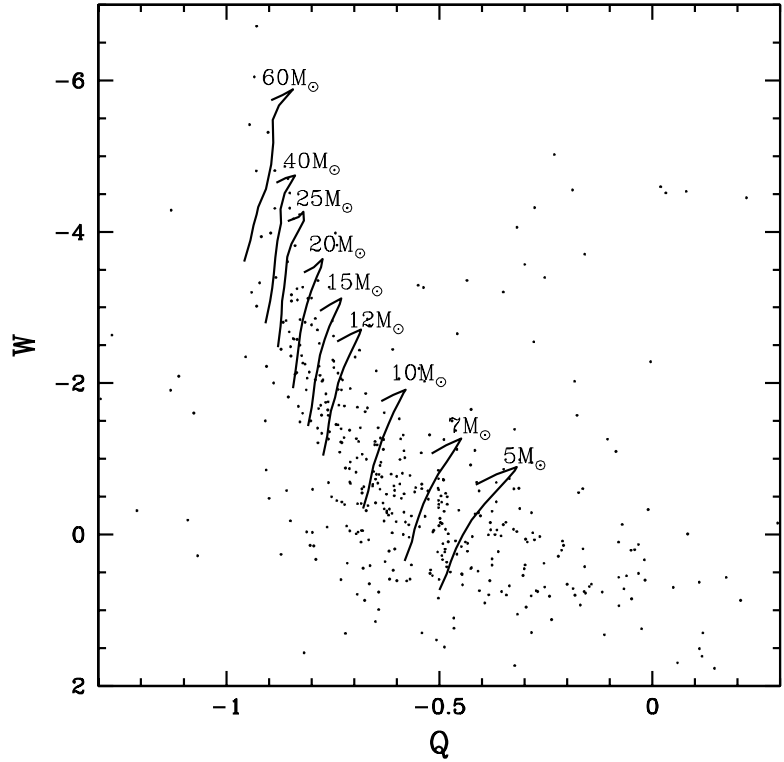}
\caption{Reddening-free color-magnitude diagram of stars in LH\,51 and
LH\,54.  The photometric data of these 456 stars are from \citet{Oey96a}.
See the text for the definition of the reddening-free magnitude W 
and color Q.  The evolutionary tracks of stars are from \citet{LS01}.
}
\label{fig4}
\end{figure}

\begin{figure}
\plotone{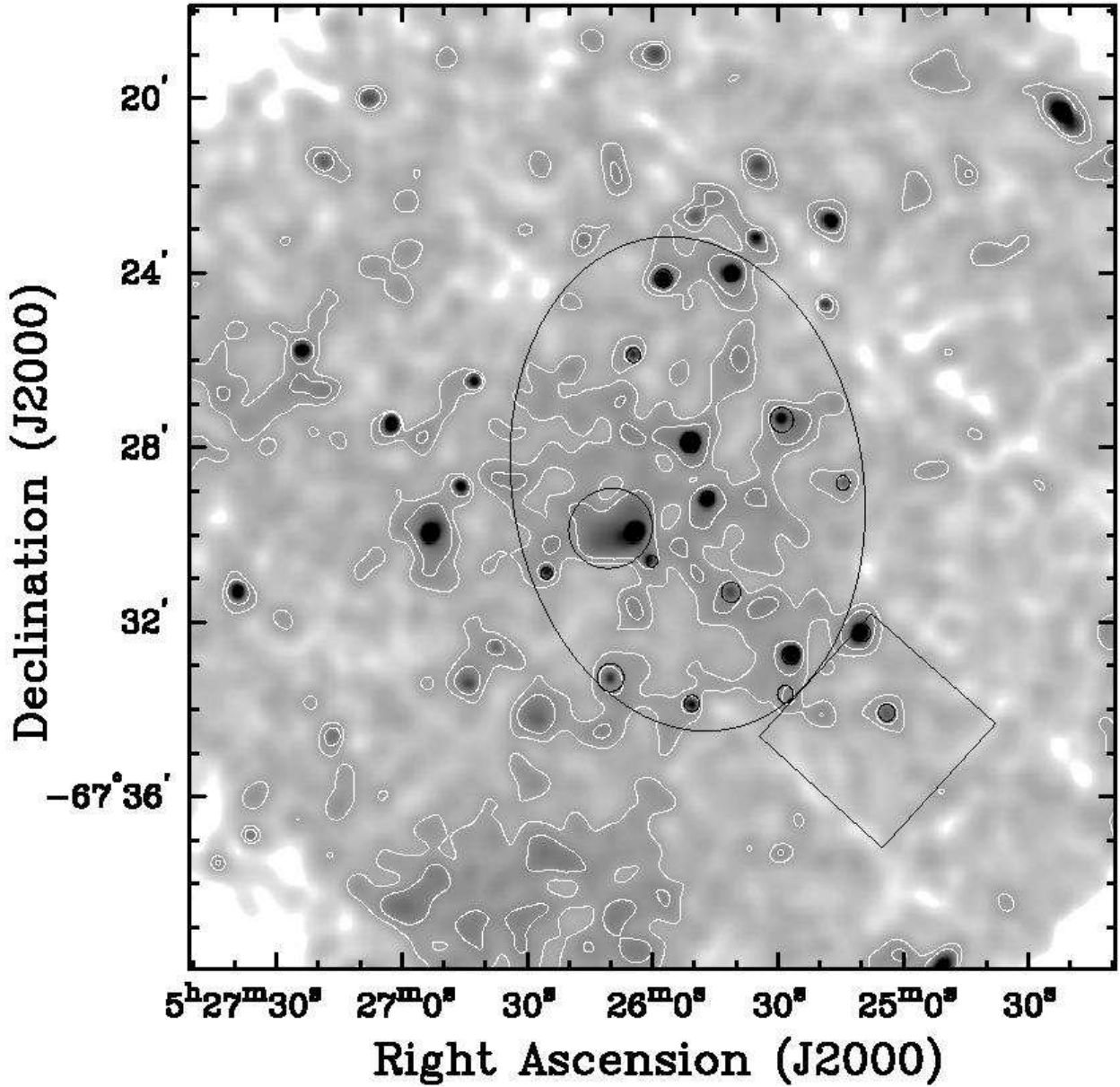}
\caption{{\sl XMM-Newton} pn image of DEM\,L\,192 in the 1.1--3.5 keV
energy band.  The image has been adaptively smoothed with Gaussians
of FWHM of 2$''$ to 15$''$.  The emission in the 1.1--3.5 keV range
is dominated by the power-law component.  The contours are at 3$\sigma$ 
and 6$\sigma$ above the background.  It is evident that the 
diffuse emission in this energy range is widespread.
The source regions are marked for comparison with Fig.~1a.
}
\label{fig5}
\end{figure}

\newpage

\begin{deluxetable}{llcccc}
\tablewidth{0pt}
\tablecaption{Stellar Data}
\tablehead{
\colhead{} & 
\colhead{} & 
\colhead{$M$} & 
\colhead{$\log {\dot{M}}$} & 
\colhead{${v_{\infty}}$} &
\colhead{$\log {L_{\rm w}}$} \\
\colhead{Star ID} & 
\colhead{Spectral Type} & 
\colhead{(M$_{\odot}$)} & 
\colhead{(M$_{\odot}$~yr$^{-1}$)} & 
\colhead{(km~s$^{-1}$)} &
\colhead{(ergs~s$^{-1}$)}
}
\startdata
L54SA-1b & O8 Iaf         & 51 & -5.170 & 1699 & 36.79\\
L54SA-2  & O8 III((f))    & 55 & -5.491 & 2064 & 36.64\\
L54S-2   & O9 Ib(f)       & 51 & -5.170 & 1699 & 36.79\\
L51N-1   & O8.5 III((f))  & 55 & -5.491 & 2064 & 36.64\\
L54S-1a  & O7: III(f)     & 55 & -5.491 & 2064 & 36.64\\
L54S-4   & O4 III(f*)     & 96 & -5.105 & 2445 & 37.17\\
L54S-5   & O8 III((f))    & 55 & -5.491 & 2064 & 36.64\\
L54S-3a  & O6.5: V        & 32 & -6.527 & 2422 & 35.54\\
L54S-3b  & B2: II:        & 19 & -6.623 & 1561 & 35.26\\
L54S-1b  & B0.5: III:     & 19 & -6.623 & 1561 & 35.26\\
L54N-2   & O6.5 V         & 32 & -6.527 & 2422 & 35.54\\
L54N-3   & B1 IIIe        & 19 & -6.623 & 1561 & 35.26\\
L54N-4   & O9 V           & 20 & -7.096 & 2205 & 35.09\\
L54S-8   & B0.5 Ib        & 34 & -5.529 & 1438 & 36.29\\
L51N-3   & O9 V           & 20 & -7.096 & 2205 & 35.09\\ 
L54S-9   & B0 III         & 19 & -6.623 & 1561 & 35.26\\
L54N-6   & O9 V           & 20 & -7.096 & 2205 & 35.09\\ 
L54S-10  & early B:: +neb & 19 & -6.873 & 1909 & 35.19\\
L54N-7   & B0.5:: V +neb  & 19 & -6.873 & 1909 & 35.19\\
L51S-2   & O6 V((f))e     & 39 & -6.409 & 2332 & 35.83\\
L54SA-4  & B0.5 III       & 19 & -6.623 & 1561 & 35.26\\
L54S-12  & B0 V \#        & 19 & -6.873 & 1909 & 35.19\\
L54S-13  & B0.5 Ve        & 19 & -6.873 & 1909 & 35.19\\
L54S-14  & O9.5 V         & 20 & -7.096 & 2205 & 35.09\\
L54S-15  & O8 Ve          & 24 & -6.835 & 2244 & 35.37\\
L54S-16  & O9.5 V \#      & 20 & -7.096 & 2205 & 35.09\\
L51S-3   & O8 V           & 24 & -6.835 & 2244 & 35.37\\
L51S-4   & B0.5 V         & 19 & -6.873 & 1909 & 35.19\\
L54S-18  & O9 V           & 20 & -7.096 & 2205 & 35.09\\
L54S-19  & B0.5: V \#     & 19 & -6.873 & 1909 & 35.19\\
L51S-5   & B0.5 V         & 19 & -6.873 & 1909 & 35.19\\
L51N-9   & B1: V +neb     & 19 & -6.873 & 1909 & 35.19\\
L51N-10  & B0 V           & 19 & -6.873 & 1909 & 35.19\\
L54S-20  & O9.5 V         & 20 & -7.096 & 2205 & 35.09\\
L54SA-5  & B1:: V +neb    & 19 & -6.873 & 1909 & 35.19\\
L51S-7   & O9.5 V         & 20 & -7.096 & 2205 & 35.09\\
L54SA-1a & WC 5           &\nodata&\nodata&\nodata&\nodata\\
L51N-12  & B0.5 V +neb    & 19 & -6.873 & 1909 & 35.19\\
L54S-22  & O9.5-B0 V      & 19 & -6.873 & 1909 & 35.19\\
L54S-24  & O9.5 V +neb    & 20 & -7.096 & 2205 & 35.09\\
L54N-14  & O9.5 V         & 20 & -7.096 & 2205 & 35.09\\
L54S-25  & B1.5 III +neb  & 19 & -6.873 & 1909 & 35.19\\
L54N-16  & O9.5 V         & 20 & -7.096 & 2205 & 35.09\\
L54S-26  & B1:: V +neb    & 19 & -6.873 & 1909 & 35.19\\
L54N-18  & O6: Ve         & 39 & -6.409 & 2332 & 35.83\\
L51S-9   & B0:: V         & 19 & -6.873 & 1909 & 35.19\\
L54SA-7  & B1.5 V         & 19 & -6.873 & 1909 & 35.19\\
L54N-19  & O9.5 V         & 20 & -7.096 & 2205 & 35.09\\
L54N-21  & B2.5 V         & 19 & -6.873 & 1909 & 35.19\\
L54S-36  & B1.5:: Ve      & 19 & -6.873 & 1909 & 35.19\\
L54N-25  & B2:: V +neb    & 19 & -6.873 & 1909 & 35.19\\
L54N-27  & O9.5-early Be  & 19 & -6.873 & 1909 & 35.19\\
L51S-14  & B1:: V         & 19 & -6.873 & 1909 & 35.19\\
L54N-33  & B1:: V         & 19 & -6.873 & 1909 & 35.19\\
\tablecomments{\# indicates spectral binary candidate.  See Table 1 of 
Oey \& Smedley (1998)}

\enddata
\end{deluxetable}

\begin{deluxetable}{lc}
\tablewidth{0pt}
\tablecaption{Energy Budget of the Superbubble DEM\,L\,192}
\tablehead{
\colhead{Energies} & 
\colhead{Amount} \\
\colhead{} &
\colhead{($\times10^{51}$ ergs)}
}
\startdata
$E_{\rm th}$ of the hot gas in superbubble interior  &  $1.1\pm0.5$ \\
$E_{\rm kin}$ of the ionized (\ion{H}{2}) superbubble shell &  $1.5\pm0.5$ \\
$E_{\rm kin}$ of the neutral (\ion{H}{1}) superbubble shell &  $3.2\pm0.5$ \\
Total energy observed in the superbubble     &  ~$6\pm2$ \\
\tableline
Stellar wind energy input in 3 Myr  &  ~$5\pm1$ \\
Supernova energy input              &  $13\pm4$ \\
Total stellar energy input          &  $18\pm5$ \\
\enddata
\end{deluxetable}

\end{document}